\documentclass[10pt,aps,prd,floatfix,notitlepage,showpacs,nofootinbib,superscriptaddress]{revtex4-1}

\usepackage{latexsym}
\usepackage{epsfig}
\usepackage{epstopdf,float}
\usepackage{graphicx}
\usepackage{amssymb}
\usepackage{amsmath}
\usepackage{dcolumn}
\usepackage{bm}
\usepackage{color}
\usepackage{comment}
\usepackage{titlesec}

\titleformat{\paragraph}
{\normalfont\normalsize\bfseries}{\theparagraph}{1em}{}
\titlespacing*{\paragraph}
{0pt}{3.25ex plus 1ex minus .2ex}{1.5ex plus .2ex}

\graphicspath{{pictures/}}

\begin{document}
\title{\bf Generalized Second Law of Thermodynamic in Modified Teleparallel Theory}

\author{M. Zubair}
\email{mzubairkk@gmail.com; drmzubair@ciitlahore.edu.pk}
\affiliation{Department of Mathematics, COMSATS Institute of Information Technology Lahore, Pakistan}

\author{Sebastian Bahamonde}
\email{sebastian.beltran.14@ucl.ac.uk}
\affiliation{Department of Mathematics, University College London, Gower Street, London, WC1E 6BT, UK}

\author{Mubasher Jamil}
\email{mjamil@sns.nust.edu.pk}
\affiliation{Department of Mathematics, School of Natural
	Sciences (SNS), National University of Sciences and Technology
	(NUST), H-12, Islamabad, Pakistan}

\date{\today}

\begin{abstract}
	This study is conducted to examine the validity of generalized
	second law of thermodynamics (GSLT) in flat FRW for modified teleparallel gravity
	involving coupling between a scalar field with the torsion scalar $T$
	and the boundary term $B=2\nabla_{\mu}T^{\mu}$. This theory is very useful since it can
	reproduce other important well-known scalar field theories in suitable limits. The validity of the first and second law of thermodynamics at the apparent horizon  is discussed for any coupling. As examples, we have also explored the validity of those thermodynamics laws in some new cosmological solutions under the theory.  Additionally, we have also
	considered the logarithmic entropy corrected relation and discuss
	the GSLT at the apparent horizon. 
\end{abstract}

\maketitle 

\section{Introduction}

The rapid growth of observational measurements on expansion history
reveals the expanding paradigm of the universe. This fact is based
on accumulative observational evidences mainly from Type Ia
supernova and other renowned sources \cite{1}. The expanding phase
implicates the presence of a repulsive force which compensates the
attractiveness property of gravity on cosmological scales. This
phenomenon may be translated as the existence of exotic matter
components and most acceptable understanding for such enigma is
termed as dark energy (DE) having large negative pressure. Various
DE models and modified theories of gravity have been proposed to
incorporate the role of DE in cosmic expansion history (for review
see \cite{2}).

In contrast to Einstein's relativity and its proposed modifications
where the source of gravity is determined by curvature scalar terms,
another formulation is presented which comprises torsional
formulation as gravity source \cite{3}. This theory is labeled as
TEGR (teleparallel equivalent of general relativity) and it is
determined by a Lagrangian density involving a zero curvature
Weitzenb$\ddot{o}$ck connection instead of a zero torsion
Levi-Civita connection with the vierbein as a fundamental tool. The
Weitzenb$\ddot{o}$ck connection is an specific connection which
characterizes a globally flat space-time endowed with a non-zero
torsion tensor. Using that connection, one can construct an
alternative and equivalent theory of GR. The later comes out since
the scalar torsion only differs by a boundary term
$B=\frac{2}{e}\partial_\mu(eT^\mu)$ with the scalar curvature by the
relationship $R=-T+B$, making both variations of the
Einstein-Hilbert and TEGR actions the same. Thus, these two theories have the same field equations. However, these two theories have
different geometrical interpretations since in TEGR the torsion acts
a force meanwhile in GR, the gravitational effects are understood due to the
curved space-time. TEGR is then further extended to a
generalized form by the inclusion of a $f(T)$ function in the
Lagrangian density (as $f(R)$ is the extension of GR) and it has
been tested cosmologically by numerous researchers \cite{4}. It is
important to mention that $f(T)$ and $f(R)$ are no longer equivalent
theories, and in order to consider the equivalent teleparallel
theory of $f(R)$, one needs to consider a more generalized function
$f(T,B)$, incorporating the boundary term in the action
\cite{Bahamonde:2015zma}. In \cite{Bahamonde:2016cul}, the authors studied some cosmological features (reconstruction method, thermodynamics and stability) within $f(T,B)$ gravity and in \cite{Bahamonde:2016grb}, some cosmological solutions were found using the Noether's symmetry approach. Additionally, it has been proved that when one consider Gauss-Bonnet higher order terms, an additional boundary term $B_{G}$ (related to the contorsion tensor) needs to be introduced to find the equivalent teleparallel modified Gauss-Bonnet theory $f(R,G)$ \cite{Bahamonde:2016kba}.

Later, Harko et al. \cite{5} proposed a comprehensive form of this
theory by involving a non-minimal torsion matter interaction in the
Lagrangian density. In a recent paper \cite{6}, Zubair and Waheed
have investigated the validity of energy constraints for some
specific $f(T)$ models and discussed the feasible bounds of involved
arbitrary parameters. They also discussed the validity of
generalized second law of thermodynamic in cosmological constant
regime \cite{6*}.

Another very studied approach in modified theories of gravity is to change
the matter content of the universe by adding a scalar field in the matter
sector. These models have been considered several times in cosmology, having
different kind of scalar fields such as quintessence, quintom, k-essence, etc
(See \cite{Copeland:2006wr,Cai:2009zp,ArmendarizPicon:2000ah}). Moreover, we
can also extend that idea by adding a coupling between the scalar field and
the gravitational sector (See \cite{17,Perrotta:1999am,Bertolami:1999dp})
where cosmologically speaking, we can have new interesting results such as
the possibility of crossing the phantom barrier. Motivated with these
theories, another interesting modified theories of gravity have also been
discussed in the literature \cite{7} on cosmological landscape. Recently,
Bahamonde and Wright \cite{8} presented a new model of teleparallel gravity
by introducing a scalar field non-minimally coupled to both the torsion $T$
and the boundary term $B=2\nabla_{\mu}T^{\mu}$. It is shown that such theory can describe the
non-minimal coupling to torsion and also non-minimal coupling to scalar
curvature under certain limits.

Black hole thermodynamics suggests that there is a fundamental
connection between gravitation and thermodynamics \cite{9}. Hawking
radiation \cite{10}, the proportionality relation between the
temperature and surface gravity, and also the connection between
horizon entropy and the area of a black hole \cite{11} support this
idea. Jacobson \cite{12} was the first to deduce the Einstein field
equations from the Clausius relation
$T_hd\hat{S}_h={\delta}\hat{Q}$, together with the fact that the
entropy is proportional to the horizon area. In the case of a
general spherically symmetric spacetime, it was shown that the field
equations can be constituted as the first law of thermodynamics
(FLT) \cite{13}. The relation between the FRW equations and the FLT
was shown in \cite{14} for $T_h=1/2{\pi}R_A,~ S_h=\pi R_A^2/G$. The
investigation about the validity of thermodynamical laws in GR as
well as modified theories has been carried out by numerous
researchers in the literature \cite{15}.

In this work we are focused on the validity of thermodynamical laws
in a modified teleparallel gravity involving a non-minimal coupling
between both torsion scalar and the boundary term with a scalar
field. The present paper is coordinated in this format. In Sec.
II, we give a brief introduction of this theory and then we derive
the respective field equation for flat FRW geometry with a perfect fluid
as the matter contents. In
Sec. III and IV, we formulate the first and second law of thermodynamics  at the apparent horizon for any coupling. Additionally, as examples, we
study the validity of the thermodynamics laws using new cosmological solutions based on power-law and
exponential law cosmology (see appendix~\ref{appendix}). In Sec. V, we discuss the validity of GSLT for
the entropy functional with quantum corrections. In each case,
suitable limits of the parameters are chosen in order to visualize
the validity of GSLT in quintessence, scalar field nonminimally
coupled to torsion (known as teleparallel dark energy) and
non-minimally coupled to the scalar curvature theories. Finally, in
Sec. VI, a discussion of the work is presented. 

In the following paper, the notation used is the same as in \cite{8}, where
the tetrad and the inverse of the tetrad fields are denoted by a lower letter
$e^{a}_{\mu}$ and a capital letter $E^{\mu}_{a}$ respectively with the $(+,-,-,-)$ metric signature.

\section{Teleparallel quintessence with a nonminimal coupling to a boundary term}

In this study we consider the modified teleparallel model which
involves scalar field non-minimally coupled to torsion $T$ and a
boundary term defined in terms of divergence of torsion vector
$B=\frac{2}{e}\partial_\mu(eT^\mu)$, where $e=\textrm{det}(e^{a}_{\mu})$. The action of this theory is
given by

\begin{align}
    S = \int
    \left[
    \frac{1}{2\kappa^2}( f(\phi) T +g(\phi) B)+\frac{1}{2}\partial_\mu \phi \partial^\mu \phi -V(\phi) +L_{\rm m}\right] e\, d^4x,\label{1}
\end{align}
where $L_{m}$ determines the matter contents, $\kappa^2=8\pi G$, $V(\phi)$ is the energy potential and $f(\phi)$ and $g(\phi)$ are coupling functions. For
simplicity, we will use the notation NMC-(B+T) to refer to this
theory  $(f(\phi)\neq g(\phi)\neq0)$. In Ref. \cite{8}, the
authors consider a special case of this action where
$f(\phi)=1+\kappa^2\xi\phi^2$ and $g(\phi)=\kappa^2\chi\phi^2$.  Non-minimally coupled scalar
field with the boundary term $B$ (NMC-B) is recovered if $\xi=0$.
Using dynamical system techniques, the cosmology in NMC-B was
studied in \cite{8}. If $\chi=0$, we get the same action as in
\cite{17}, which is known as ``teleparallel dark energy theory"
(TDE). By choosing $\chi=-\xi$ we obtain scalar field models
non-minimally coupled to the Ricci scalar that hereafter, for
simplicity we will label as NMC-R \cite{Perrotta:1999am}. In
addition, the so-called ``Minimally coupled quintessence" theories
arise when we take $\chi=\xi=0$ \cite{Bertolami:1999dp}.

Variation of the action (\ref{1}) with respect to the tetrad field
yields the following field equations
\begin{align}
\frac{2}{\kappa^2}f(\phi)\left[ e^{-1}\partial_\mu (e S_{a}{}^{\mu\nu})-E_{a}^{\lambda}T^{\rho}{}_{\mu\lambda}S_{\rho}{}^{\nu\mu}-\frac{1}{4}E^{\nu}_{a}T\right]-E^{\nu}_a \left[\frac{1}{2}\partial_\mu \phi \partial^\mu \phi -V(\phi)\right]
\nonumber\\ +E^{\mu}_a \partial^\nu \phi \partial_\mu \phi + \frac{1}{\kappa^2}\Big[2(\partial_{\mu}f(\phi)+\partial_{\mu}g(\phi)) E^\rho_a S_{\rho}{}^{\mu\nu}+E^{\nu}_{a}\Box g(\phi)-E^\mu_a \nabla^{\nu}\nabla_{\mu}g(\phi)\Big]=  \mathcal{T}^\nu_a, \label{2}
\end{align}
where ${\Box}={\nabla}_{\alpha}{\nabla}^{\alpha};~{\nabla}_{\alpha}$ is the
covariant derivative linked with the Levi-Civita connection symbol and
$\mathcal{T}^{\nu}_a$ is the matter energy momentum tensor.\\
If we vary the action (\ref{1}) with respect to the scalar field, yields the following modified
Klein-Gordon equation
\begin{eqnarray}
\Box \phi+V'(\phi)&=\frac{1}{2\kappa^2}\Big(f'(\phi)T+g'(\phi)B\Big).\label{5}
\end{eqnarray}
Throughout this paper, prime denotes differentiation with respect to $\phi$. We
will assume the homogeneous and isotropic flat FRW metric in
Euclidean coordinates defined as
\begin{equation}\label{6}
ds^2=dt^2-a^2(t)(dx^2+dy^2+dz^2)\,,
\end{equation}
where $a(t)$ represents the scale factor and the corresponding tetrad components
are $e^i_\mu=(1,a(t),a(t),a(t))$. It is important to mention that in despite that $f(T)$ gravity is not invariant under local Lorentz transformations, the later tetrad is a ``good tetrad" to consider since it does not constraint its field equations \cite{Tamanini:2012hg}. Hence, this tetrad can be used safely within this scalar field theory too. \\
The energy-momentum tensor of matter is defined as
\begin{equation}\label{7}
\mathcal{T}_{{\mu}{\nu}}=({\rho}_m+p_m)u_{\mu}u_{\nu}-p_{m}g_{{\mu}{\nu}},
\end{equation}
where $u_{\mu}$ is the four velocity of the fluid and $\rho_m$ and $p_m$ define the
matter energy density and pressure respectively.
Using the tetrad components for the flat FRW metric, the field equations (\ref{2})
lead to
\begin{align}
3H^{2}f(\phi)&=\kappa^2\Big(\rho_m+V(\phi)+\frac{1}{2}\dot{\phi}^{2}\Big)+3H\dot{\phi}g'(\phi),\label{FE1}\\
3H^2f(\phi)+2\dot{H}f(\phi)&=-\kappa^2\Big( p_m-V(\phi)+\frac{1}{2}\dot{\phi}^{2}\Big)-2H\dot{\phi}f'(\phi)+
(g''(\phi)\dot\phi^2+g'(\phi)\ddot \phi).\label{FE2}
\end{align}
Here, $H=\dot{a}(t)/a(t)$ is the Hubble parameter and dots and primes denote differentiation with respect to the time coordinate and the argument of the function respectively. We
can also rewrite these equations in a fluid representation,
\begin{eqnarray}\label{8}
3H^2&=&\kappa_\text{eff}^2\rho_\text{eff},\\\label{9}
2\dot{H}&=&-\kappa_\text{eff}^2(\rho_\text{eff}+p_\text{eff}),
\end{eqnarray}
where $\kappa^2_\text{eff}=\frac{\kappa^2}{f(\phi)}$,
$\rho_\text{eff}=\rho_{m}+\rho_\phi$ is the total energy density and
$p_\text{eff}=p_{m}+p_\phi$ is the total pressure. The energy
density and the pressure of the scalar field $\rho_\phi$ and
$p_\phi$ are respectively defined as follows
    \begin{eqnarray}
    \rho_\phi&=&\frac{1}{2}\dot{\phi}^2+V(\phi)+\frac{3}{\kappa^2} H\dot{\phi}g'(\phi)\,,\nonumber\\
    p_\phi&=&\frac{1}{2}\dot{\phi}^2-V(\phi)+\frac{1}{\kappa^2}\Big(2H\dot{\phi}f'(\phi)-(g''(\phi)
    \dot\phi^2+g'(\phi)\ddot \phi)\Big)\,.\label{pp}
    \end{eqnarray}
In this theory the standard continuity equation reads
\begin{eqnarray}\label{13}
\dot{\rho}_\text{eff}+3H(\rho_\text{eff}+p_\text{eff})=0\,,\\
\dot{\rho}_{m}+3H(\rho_{m}+p_{m})=0\,.
\end{eqnarray}
Hereafter, we will assume a standard equation of state for the
matter given by a barotropic equation $p_{m}=w \rho_{m}$, with $w$
being the state parameter. If we use the above equation, we can
directly find that the energy density becomes
\begin{eqnarray}\label{phi}
\rho_{m}&=&\rho_{0}a(t)^{-3(1+w)}\,,
\end{eqnarray}
where $\rho_{0}$ is an integration constant. It is proper to mention
that for a flat FRW metric, the torsion scalar and the boundary term
are
\begin{align}
    T&=-6H^2\,,\\
     B&=-18H^2-6\dot{H}\,,
    \end{align}
and hence the Ricci scalar is recovered via $R=-T+B=-12H^2-6\dot{H}$.\\
Finally, the equation for the scalar field, the so-called Klein-Gordon equation takes the form
\begin{align}
\ddot{\phi}+3H\dot{\phi}-\frac{1}{2\kappa^2}\Big(f'(\phi)T+g'(\phi)B\Big)+V'(\phi)&=0.\label{12}
\end{align}
Note that the Klein-Gordon equation can be also obtained directly from the field
equations (\ref{FE1}) and (\ref{FE2}), so that it is not an extra equation.

\section{Thermodynamics in Modified Teleparallel Theory}

In this section we are interested to explore the general thermodynamic laws in the
framework of the theory studied in \cite{8}, where the authors
considered a quintessence theory non-minimally coupled between both
a torsion scalar $T$ and the boundary term $B$ with the scalar field (NMC-B+T). The main aim of the next sections will be to formulate the first and second laws of thermodynamics in this theory. The complete and general thermodynamics law will be derived. After that, we will use the cosmological solutions found in the appendix~\ref{appendix} to study some interesting examples to visualize if them satisfy or not the thermodynamic laws.

\subsection{First Law of Thermodynamics}
This section is devoted to investigate the validity of the first law of thermodynamics in
NMC-(B+T) at the apparent horizon for a flat FRW universe.\\
The condition $h^{\alpha\beta}\partial_{\alpha}R_A\partial_{\beta}R_A=0$
gives the radius $R_A$ of apparent horizon for flat FRW metric as
\begin{equation*}
R_A=\frac{1}{H}.\label{RA}
\end{equation*}
The associated temperature is $T_h=\kappa_{sg}/2\pi$, where the
surface gravity $\kappa_{sg}$ is given by \cite{19}
\begin{eqnarray}\label{zz**}
\kappa_{sg}&=&\frac{1}{2\sqrt{-h}}\partial_{\alpha}(\sqrt{-h}h^
{\alpha\beta}\partial_{\beta}R_A)=-\frac{1}{R_A}
\Big(1-\frac{\dot{R}_A}{2HR_A}\Big)\nonumber\\
&=&-\frac{R_A}{2}(2H^2+\dot{H}).
\end{eqnarray}
By using Eqs (\ref{9}) and (\ref{RA}) we easily get
	\begin{equation}\label{ddr}
	\frac{f(\phi)}{G}d{R_A}= 4\pi R_A^3H({\rho}_\text{eff}+{p_\text{eff}})dt.
	\end{equation}
Now, multiplying both sides of this equation by a factor
$-2\pi R_{A}T_{h}=1-\dot{R}_A/(2HR_A)$, we can rewrite the above equation as follows
\begin{eqnarray}\label{dSTH}
T_h d\Big(\frac{Af(\phi)}{4G}\Big)&=&-(4\pi R_{A}^3Hdt-2\pi R_{A}^2dR_{A})(\rho_{eff}+p_{eff})+\frac{\pi}{G}R_{A}^2T_{h}df(\phi)\,,\label{}
\end{eqnarray}
where we have used that $A=4\pi R_{A}^2$. From here, we can identify as the entropy as the quantity which is multiplied by $T_{h}$, namely
\begin{equation}\label{entr}
S_h=\frac{Af(\phi)}{4G}\,.
\end{equation}
Now, we define energy of the universe within the apparent horizon.
The Misner-Sharp energy is defined as $E=R_A/(2G_{eff})$, which can be
written as
\begin{equation}\label{ee}
E=\frac{R_Af(\phi)}{2G}.
\end{equation}
In terms of volume $V=4\pi R_A^3/3$, we obtain that the energy density is given by
\begin{equation}\label{E}
\bar{E}=\frac{3H^2f(\phi)}{8\pi G}V\equiv{\rho}_\text{eff}V,
\end{equation}
Taking differential of energy relation, we get
\begin{equation}\label{zz}
d\bar{E}=\frac{R_A}{2G}df(\phi)+4\pi {\rho}_\text{eff} {R_A}^2d
R_A-4\pi H{R_A}^3({\rho}_\text{eff}+{p_\text{eff}})dt.
\end{equation}
Combining Eqs.~(\ref{dSTH}) and (\ref{zz}), it results in
\begin{equation}\label{E11}T_hdS_h=d\bar{E}-2\pi
{R_A}^2({\rho}_\text{eff}-{p_\text{eff}})dR_A+\frac{R_A}{2G}(2\pi R_{A}T_{h}-1)df(\phi).
\end{equation}
By defining the work density, we get
\begin{equation}\label{E1}
\bar{W}=-\frac{1}{2}\left(
T^{(M)\alpha\beta}h_{\alpha\beta}+\bar{T}^{(de)\alpha\beta}h_{\alpha\beta}\right)=\frac{1}{2}({\rho}_\text{eff}-{p_\text{eff}}).
\end{equation}
Here $\bar{T}^{(de)\alpha\beta}h_{\alpha\beta}$ is energy-density of
the dark components. Using above definition of work density in
Eq. (\ref{E11}), we obtain
\begin{equation}\label{E111}
T_hdS_h=d\bar{E}-\bar{W}dV+\frac{R_A}{2G}(2\pi R_AT_h-1)df(\phi),
\end{equation}
which can be re-written as
\begin{equation}\label{E4}
T_hdS_h+T_hdS_p=d\bar{E}-\bar{W}dV,
\end{equation}
where $dS_p=-(R_{A}/(2GT_h))(2\pi R_AT_h-1)df(\phi)$, which is the
first law of thermodynamics in this teleparallel theory. The extra term
$dS_p$ defined in Eq.~(\ref{E4}) can be treated as an entropy
production term in non-equilibrium thermodynamics. In gravitational theories such as Einstein, Gauss-Bonnet and Lovelock gravities \cite{7*}, the usual first law of thermodynamics is satisfied by the
respective field equations. In fact these theories do not involve
any surplus term in universal form of first law of thermodynamics
\emph{i.e.}, $TdS=dE-WdV$.

Initially Akbar and Cai used this approach to discuss thermodynamic laws in
$f(R)$ gravity \cite{25*}. It is shown that an additional entropy term produces as compared to other modified theories.
Later Bamba et al. \cite{26*} developed the first law of thermodynamics in Palatini $f(R)$, $f(T)$, $f(R,\phi,X)$ (where $X=-1/2g^{\mu\nu}\nabla_\mu\phi\nabla_\nu\phi$ is the kinetic term of a scalar field $\phi$) and $f(R,\phi,X,G)$ (where $G=R^2-4R_{\mu\nu}R^{\mu\nu}+R_{\mu\nu\rho\sigma}R^{\mu\nu\rho\sigma}$ is the Gauss-Bonnet
invariant) theories and formulated additional entropy production term. Similar approach is applied to discuss thermodynamic laws in $f(R,T)$, $f(R,L_m)$ and $f(R,T,R_{\mu\nu}T^{\mu\nu})$ theories, one can see that presence of non-equilibrium entropy production terms is necessary in such theories \cite{7**}. Bamba et al. \cite{26*} have shown that one can manipulate the FRW 
equations in order to redefine the entropy relation, which results in equilibrium description of thermodynamics so that first law of thermodynamics takes the form $TdS=dE-WdV$. Moreover, in all these theories it has been that usual form form of first law of thermodynamics \emph{i.e.}, $TdS_{eff}=dE-WdV$ can be obtained by defining the general entropy relation as sum of horizon entropy and entropy production term.

Here, we may define the effective
entropy term being the sum of horizon entropy and entropy production
term as $S_\text{eff}=S_h+S_p$ so that Eq.(\ref{E4}) can be
rewritten as
\begin{equation}\label{zu}
T_hdS_\text{eff}=d\bar{E}-\bar{W}dV,
\end{equation}
where $S_\text{eff}$ is the effective entropy related to the
contributions involves scalar field non-minimally coupled to torsion
$T$ and a boundary term at the apparent horizon of FRW spacetime.

\subsection{Generalized Second Law of Thermodynamics}
According to the generalised second law of thermodynamic (GSLT), the entropy of matter and energy sources inside
the horizon plus the entropy associated with boundary of horizon
must be non-decreasing. In previous section we have shown that usual first law of thermodynamics does not hold in this theory.
Therefore to study GSLT, we would use the modified first law of thermodynamics. In fact the generalized entropy relation
satisfies the condition
\begin{equation}\label{14*}
\dot{S}_{tot}=\dot{S}_h+\dot{S}_p+\dot{S}_{in}\geq0\,,
\end{equation}
where $\dot{S}_h$ represents the entropy associated with the horizon, $\dot{S}_p$ represents the entropy production term
and $\dot{S}_{in}$ is the sum of all entropy components inside the
horizon.\\
Let us proceed with the modified first law of thermodynamics
\begin{equation}\label{18}
T_idS_i=dE_i+p_idV-T_idS_p\,,
\end{equation}
which can be represented as
\begin{equation}\label{19}
T_{in}\dot{S}_{in}=(\rho_i+p_i)4\pi{R_h}^2(\dot{R}_h-HR_h)+\frac{4}{3}\pi{R}^3_hQ_i-T_i\dot{S}_p\,,
\end{equation}
where $R_h$ represent the radius of the horizon, $T_{in}$ denote
temperature for all the components inside the horizon and $Q_i$ is
interaction term for ith component. Summing up the total entropy
inside the horizon, we find
\begin{equation}\label{20}
T_{in}\dot{S}_{in}=(\rho_\text{eff}+p_\text{eff})4\pi{R_h}^2(\dot{R}_h-HR_h)-T_{in}\dot{S}_p\,.
\end{equation}
Here, $$ \sum_iQ_i=0, \quad
\sum_i(\rho_i+p_i)=\rho_\text{eff}+p_\text{eff}.$$
Now, let us study the validity of GSLT at the apparent horizon.
In de Sitter space time $R_A=1/H$ and the future event
horizon becomes the same as the Hubble horizon. The time derivative of (\ref{entr}) results in
\begin{equation}\label{24}
\dot{S}_h=\frac{\pi}{H^2G}\{\dot{\phi}f'(\phi)-\frac{2\dot{H}}{H}f(\phi)\}\,.
\end{equation}

Here, we set the thermal equilibrium with $T_{in}=T_h$, so that Eq. (\ref{20}) implies
\begin{equation}\label{26}
\dot{S}_{in}+\dot{S}_{p}=\frac{1}{T_{h}}(\rho_\text{eff}+p_\text{eff})4\pi{R_A}^2(\dot{R}_A-HR_A)\,.
\end{equation}
After simplification, we get
\begin{equation}\label{27}
\dot{S}_{in}=\frac{4\pi}{G}\frac{\dot{H}(\dot{H}+H^2)}{(\dot{H}+2H^2)H^3}\,.
\end{equation}
Hence, Eq.(\ref{14*}) implies the relation of GSLT of the form
\begin{equation}\label{29}
\dot{S}_{tot}=\frac{\pi}{GH^2}\left(\frac{4\dot{H}(\dot{H}+H^2)}{H(\dot{H}+2H^2)}+\dot{\phi}f'(\phi)-\frac{2\dot{H}}{H}f(\phi)\right)\geq0\,.
\end{equation}
In the following sections we will take a look if the GSLT is satisfied for some interesting cosmological solutions
 that can be constructed from our solutions found before (see Sec.~\ref{powerlaw}-\ref{exponential}).

\subsubsection{Specific model: Power-law solutions for $f(\phi)=1+\xi\kappa^2 \phi^2$ and $g(\phi)=\chi\kappa^2\phi^2$}

In this section we will analyse if the GSLT is valid for different power-law models. In Sec. \ref{powerlaw},
 we found some specific solutions to the specific coupling $f(\phi)=1+\xi\kappa^2 \phi^2$ and $g(\phi)=\chi\kappa^2 \phi^2$.
  Here, we will focus our study for quartic ($m=-1$) and inverse potentials ($m=2/3$). Additionally, by setting the coupling
   constants $\xi$ and $\chi$, we will also focus our study on different non-minimally coupled scalar-tensor theories.\\
\paragraph{Power-law potential with $\chi\neq\frac{1}{4}$}
For this case, the energy potential reads,
\begin{eqnarray}\nonumber
    V(\phi)&=& \-\displaystyle\frac{1}{2} \phi ^{2-2/m} \phi_{0}^{2/m}
    \left(m^2+12 m n \chi -6 n^2 \xi \right)\,, \chi\neq\frac{1}{4}\,,
    m=\tfrac{2n(2 \xi +3\chi) +2 \chi \pm\sqrt{4 (2 n \xi +(3 n+1) \chi
            )^2+ 8 n \xi  (1-4 \chi)}}{2 (4 \chi -1)} \,.
\end{eqnarray}

\begin{figure}[H]
    \centering
    \includegraphics[width=0.4\textwidth]{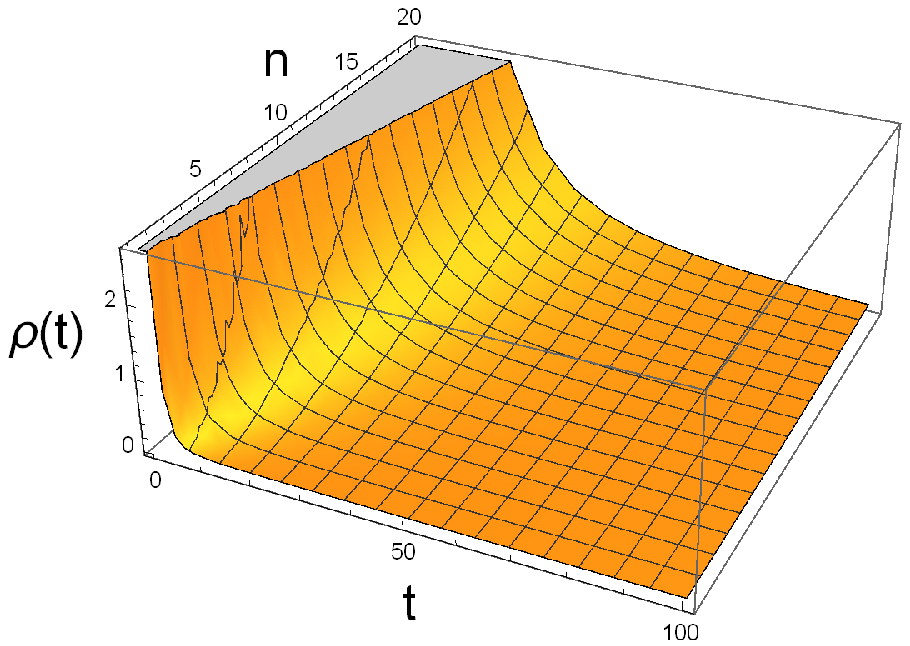}
    \includegraphics[width=0.4\textwidth]{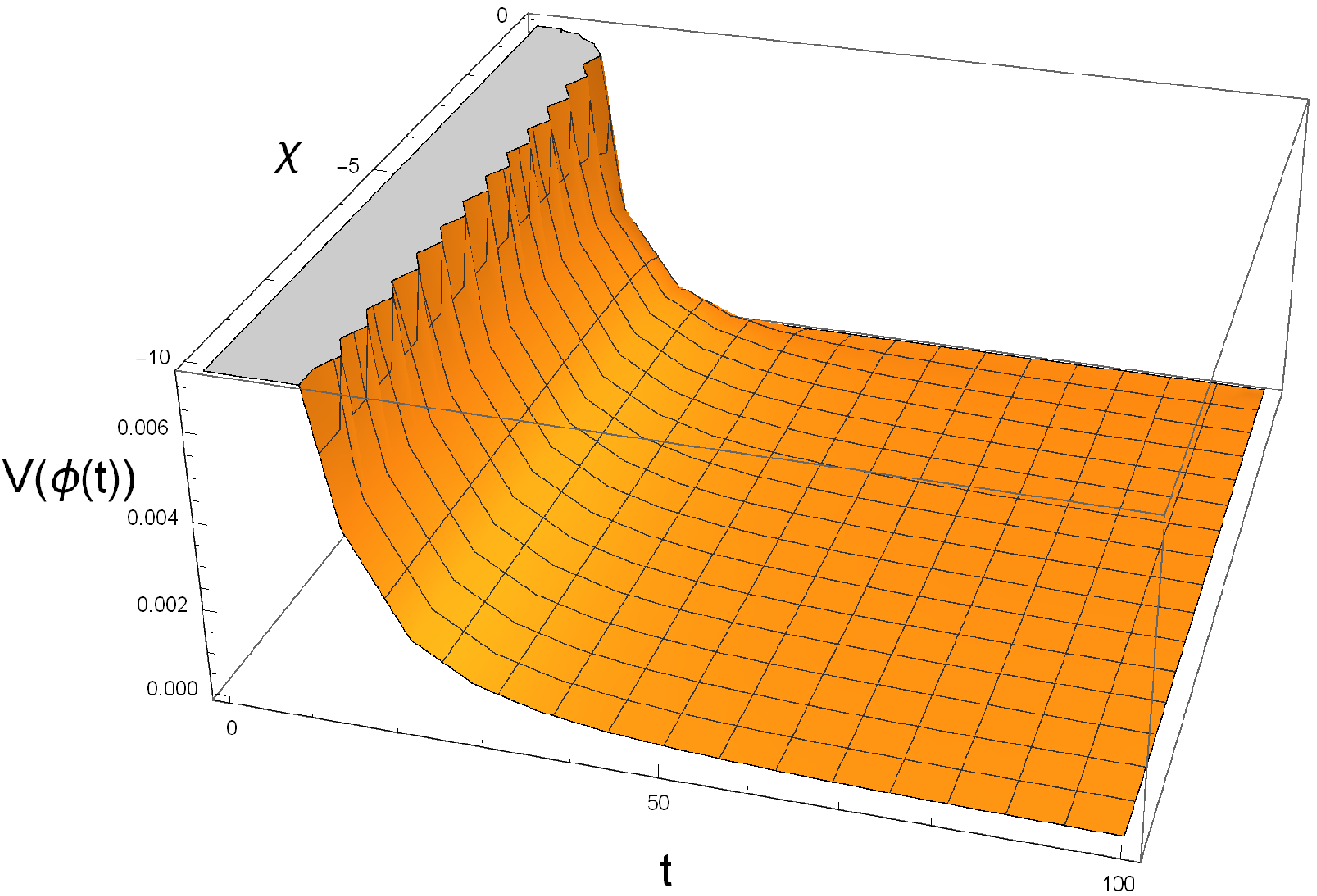}
    \caption{For the power-law potentials, the energy density and the scalar potential are plotted
        against time. We chose $\phi_{0}=\kappa=1$, $n=2$ ($a(t)\propto t^2$) and $\xi=0.1$}
\label{figG4}
\end{figure}

In Fig. \ref{figG4}, we present the evolution of energy density $\rho$ and potential $V(\phi)$ for the power law case. It can be seen that $V(\phi)$ is decreasing function of time, where we set $\xi\leq0$.

Here, we set the validity condition for GSLT in terms of parameters
$m$, $\xi$ and $\chi$. We choose $m$ to show the particular
representation of potential.

\paragraph{Case $m=-1$}

\begin{figure}[H]
    \centering
    \includegraphics[width=0.4\textwidth]{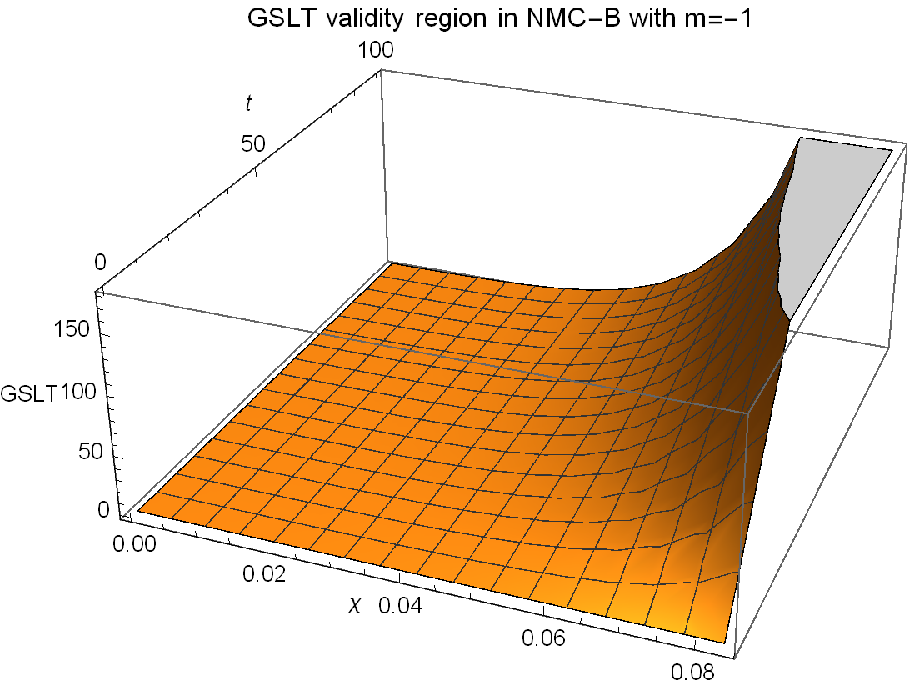}
    \includegraphics[width=0.4\textwidth]{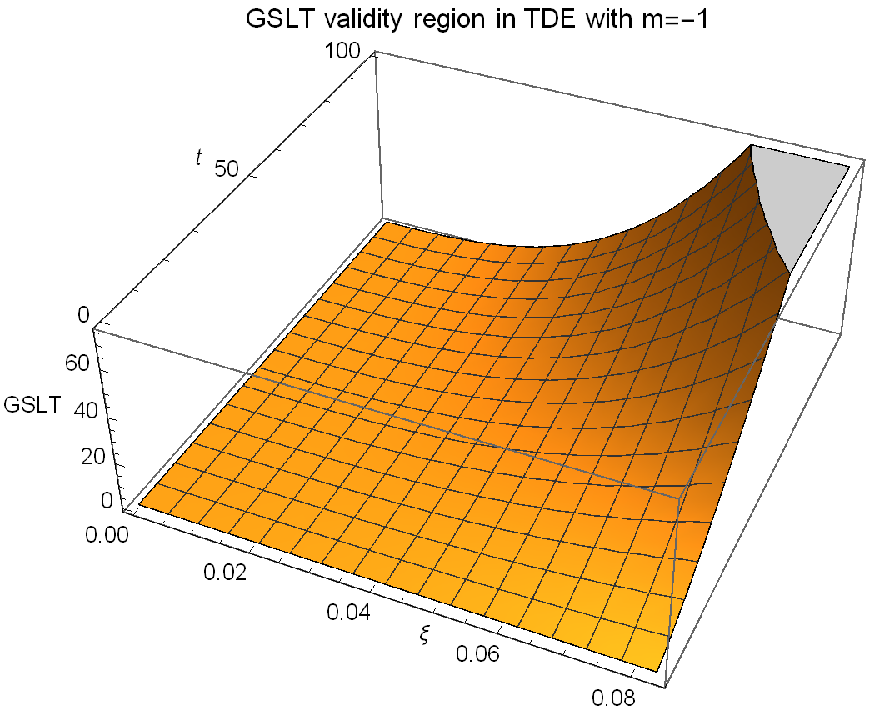}
    \caption{Validity of GSLT for power-law potential with $\chi\neq\frac{1}{4}$.
        The figure on left(right) shows the validity for NMC-B(TDE) theory.}
        \label{figP1}
\end{figure}

\begin{figure}[H]
    \centering
    \includegraphics[width=0.4\textwidth]{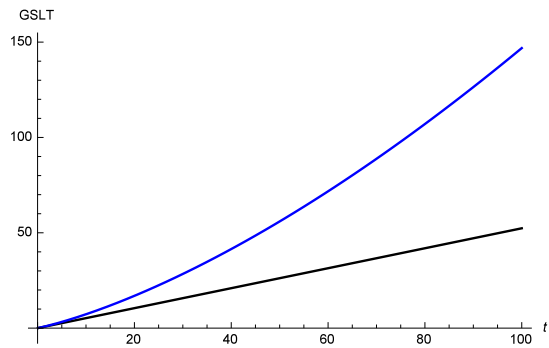}
    \caption{Validity of GSLT for power-law potential with $n=2$.}
    \label{figP2}
\end{figure}

This choice of $m$ corresponds to quartic potential
$V(\phi)\propto\phi^4$. Here, we find the relation for $n$ of the
form $n=\frac{1-6\chi}{6\xi+\chi}$ and the constraint $n>1$ results
in the following condition
\begin{equation}\label{h}
    \left(\chi < \frac{1}{6} ~\&~ -\chi< \xi<\frac{1}{6}(1 -
    12\chi)\right) || \left(\chi>\frac{1}{6} ~\&~ \frac{1}{6} (1 - 12
    \chi) < \xi < -\chi\right)
\end{equation}
Here, we discuss the specific cases NMC-B ($\xi=0$) and TDE
($\chi=0$). The validity of GSLT for quartic potential is shown in
Fig. \ref{figP1}. For NMC-B theory, the GSLT is valid in the range $0 < \chi< \frac{1}{12}$ and in case of
TDE we need $0 <\xi<\frac{1}{6}$. These constraints are set in accordance with the condition of power law solutions $n>1$.
We also the evolution of GSLT in Fig. \ref{figP2}, where we set $n=2$ and choose the particular values for $\chi$ and $\xi$.
The black curve corresponds to NMC-B with $\chi=\frac{1}{18}$ and blue curve represents the plot of GSLT for TDE with $\xi=\frac{1}{18}$.
\paragraph{Case $m=2/3$}
In this case one can recover the inverse potential i.e., $V(\phi)\propto\phi^{-1}$. Here, we show the validity of GSLT in Figs. \ref{figP3} and \ref{figP4}. In Fig. \ref{figP3}, left plot corresponds to NMC-B which shows the validity in the range $-\frac{1}{8} < \chi < 0$ and right plot corresponds to TDE with $-\frac{2}{3} < \xi < 0$. It can be seen that GSLT is violated in case of TDE whereas it is valid in both NMC-B and NMC-R.
\begin{figure}[H]\label{figP3}
    \centering
    \includegraphics[width=0.4\textwidth]{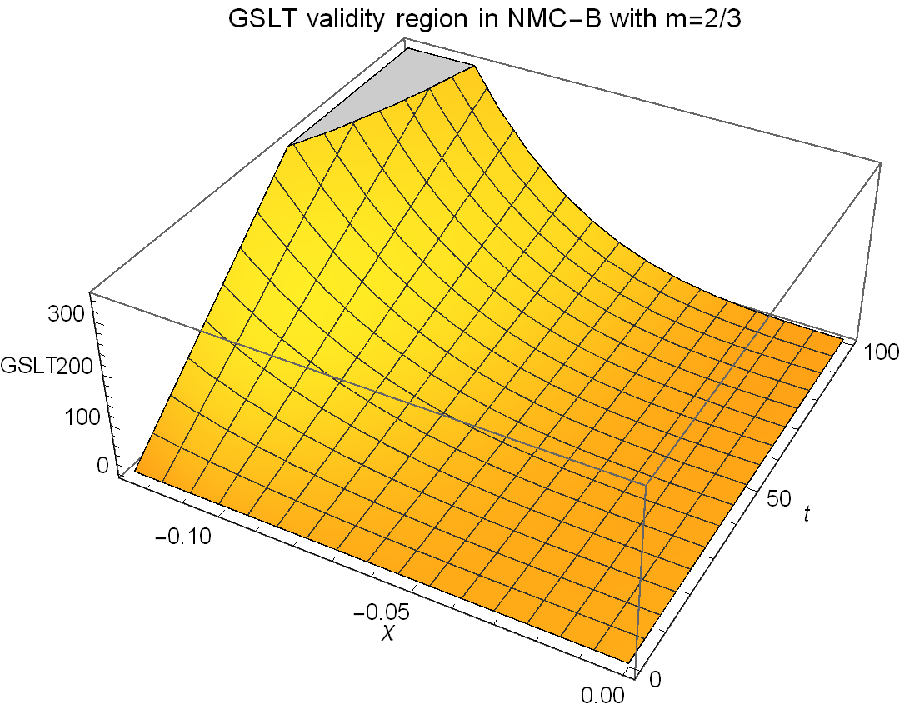}
    \includegraphics[width=0.4\textwidth]{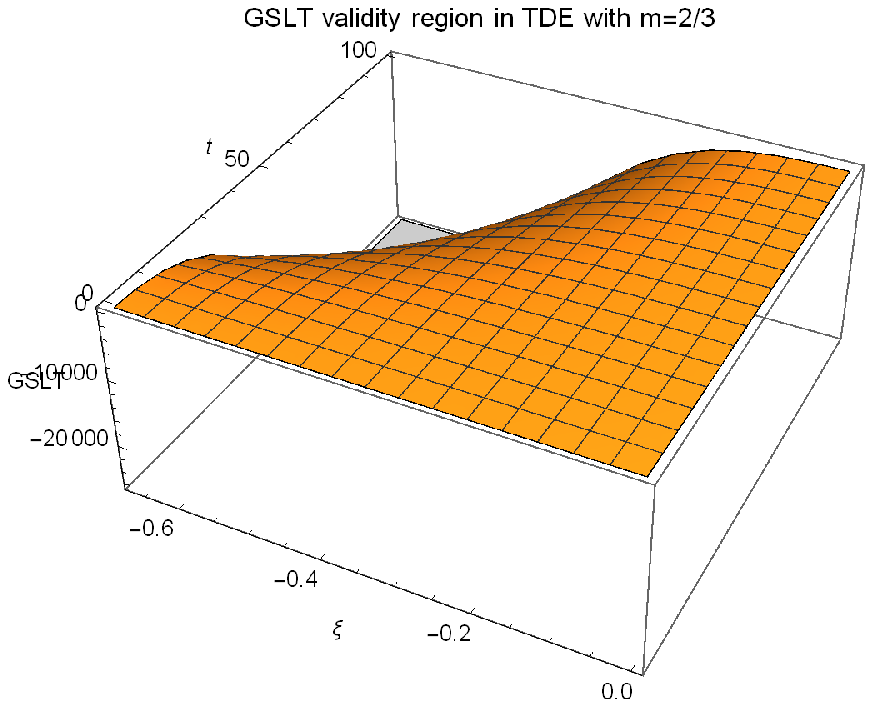}
    \caption{Validity of GSLT for power-law potential for $m=2/3$ in NMCB and TDE theories.}
    \label{figP3}
\end{figure}
\begin{figure}[H]
    \centering
    \includegraphics[width=0.4\textwidth]{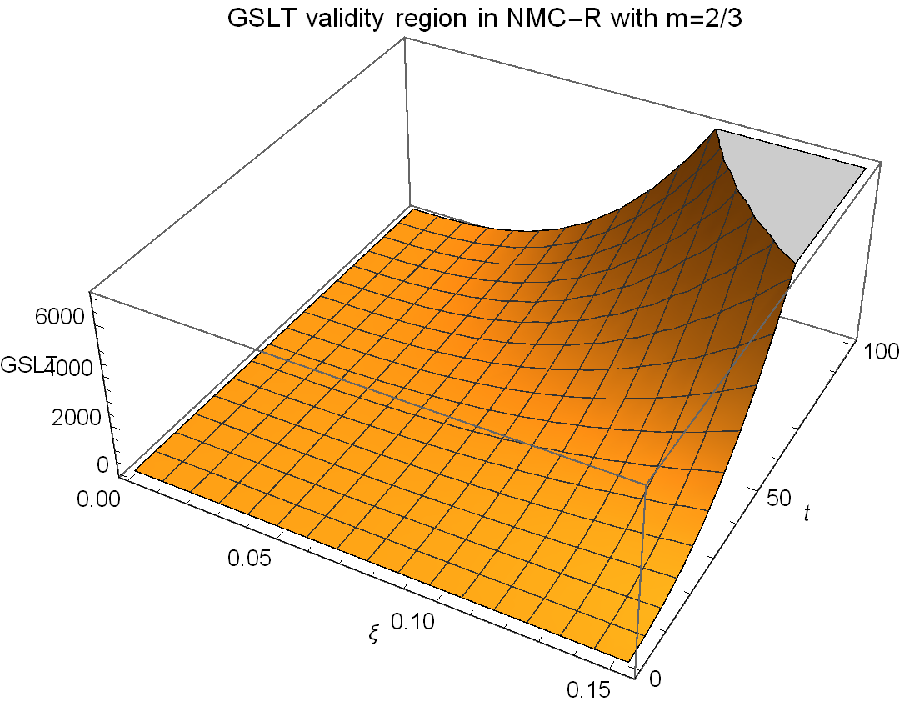}
    \includegraphics[width=0.4\textwidth]{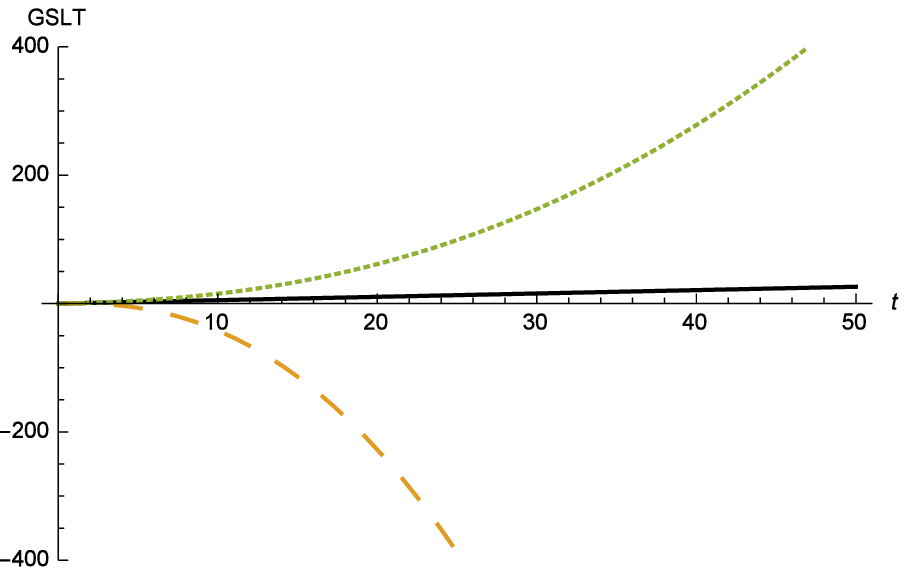}
    \caption{Left plot shows the validity of GSLT in NMC-R for $0 < \xi < \frac{2}{13}$ and right plot
        sums up the validity in case of inverse potential. The curves for NMC-B ($\chi=\frac{-1}{17}$), TDE ($\xi=\frac{-1}{3}$) and NMC-R ($\xi=\frac{1}{14}$) are represented by solid, dashed and dotted lines respectively.}
        \label{figP4}
\end{figure}
\paragraph{Power-law potential with $\chi=\frac{1}{4}$}
For this case the potential takes the following form,
\begin{eqnarray}\nonumber
    V(\phi)&=&\frac{n^2 \xi  (8 \xi +3) \left(3 n^2 (8 \xi +3)-1\right)
        \phi ^{-\frac{3 n+1}{2 n \xi }-2}\phi_{0}^{\frac{3 n+1}{2 n \xi }+4}}{(n (8 \xi +3)+1)^2}\,,
    \ \ \ \chi=\frac{1}{4}\,,m=\frac{4 n \xi }{8 \xi  n+3 n+1}\,,n\neq- \frac{1}{8 \xi +3}\,.
\end{eqnarray}
For this model, we have fixed value of $\chi$, so we can not discuss the TDE and NMC-B theories. In case of
NMC-R ($\chi=-\xi$), we find that this representation does not show realistic results as we need to fix $n<0$ for inverse and quadratic potentials.

\begin{figure}[H]
    \centering
    \includegraphics[width=0.4\textwidth]{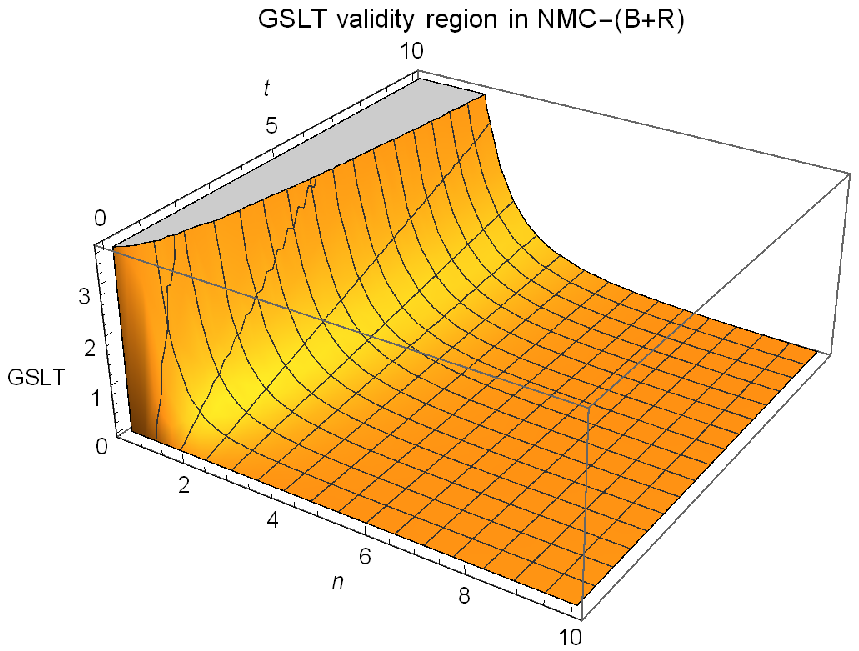}
    \includegraphics[width=0.4\textwidth]{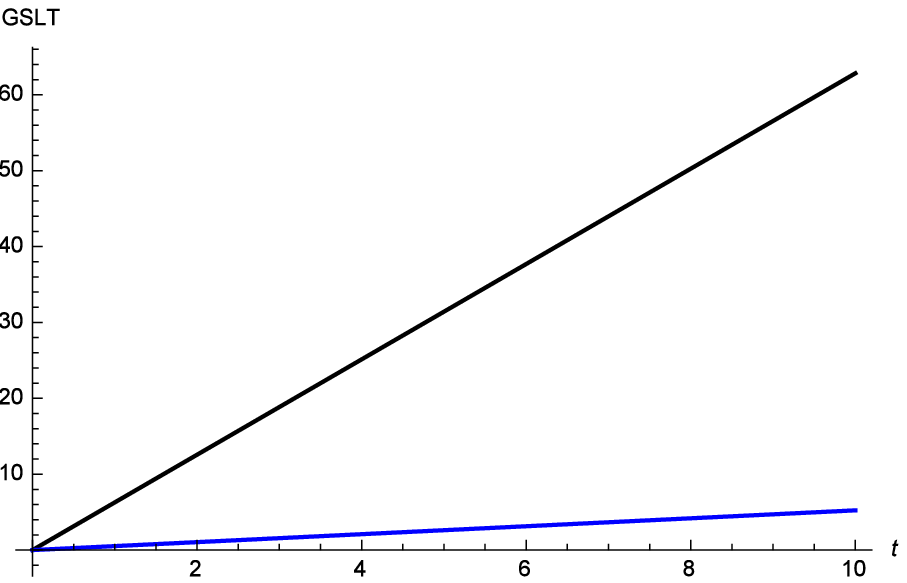}
    \caption{Here, we show the validity for NMC-(B+R) theory with quartic potential. In this case we find that GSLT
        can be met for $n>0.5$. In left plot we show the validity region, and its behavior for particular choices is shown on the right side. For black and blue curves correspond to $n=1,2$0.}
        \label{figP5}
\end{figure}

Moreover, in case of NMC-(B+R), we explore the validity for quartic potential, where we need to set $\xi=\frac{-(3n+1)}{12n}$. Finally, we have GSLT dependence on $n$, and validity is shown in Figure \ref{figP5}.

\subsubsection{Specific model: Exponential solutions for $f(\phi)=1+\kappa^2\alpha e^{m\phi}$ and $g(\phi)=\kappa^2\beta e^{k\phi}$}

\begin{figure}[H]
    \centering
    \includegraphics[width=0.4\textwidth]{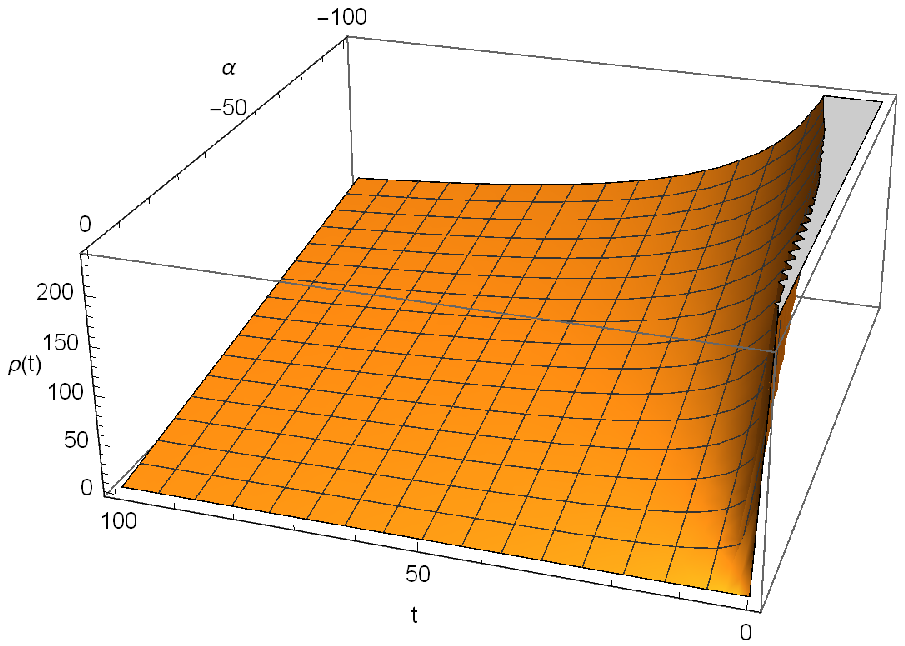}
    \includegraphics[width=0.4\textwidth]{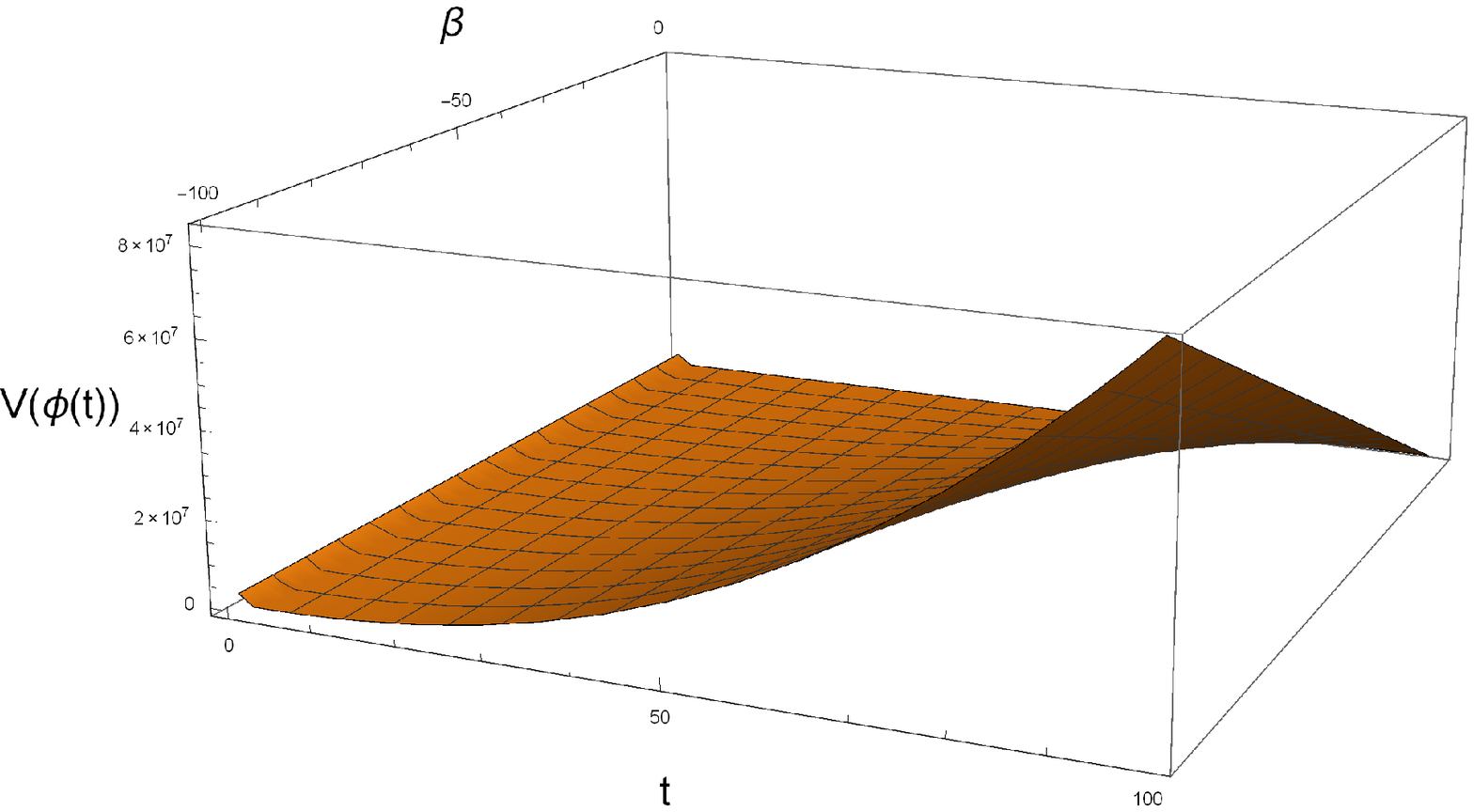}
    \caption{For the exponential law model, the energy density and the scalar potential are plotted
        against time. We chose $\phi_{0}=1.5$, $\kappa=1$, $m=1$ and $a_{0}=1$}
    \label{Figg6}
\end{figure}

\begin{figure}[H]
    \centering
    \includegraphics[width=0.4\textwidth]{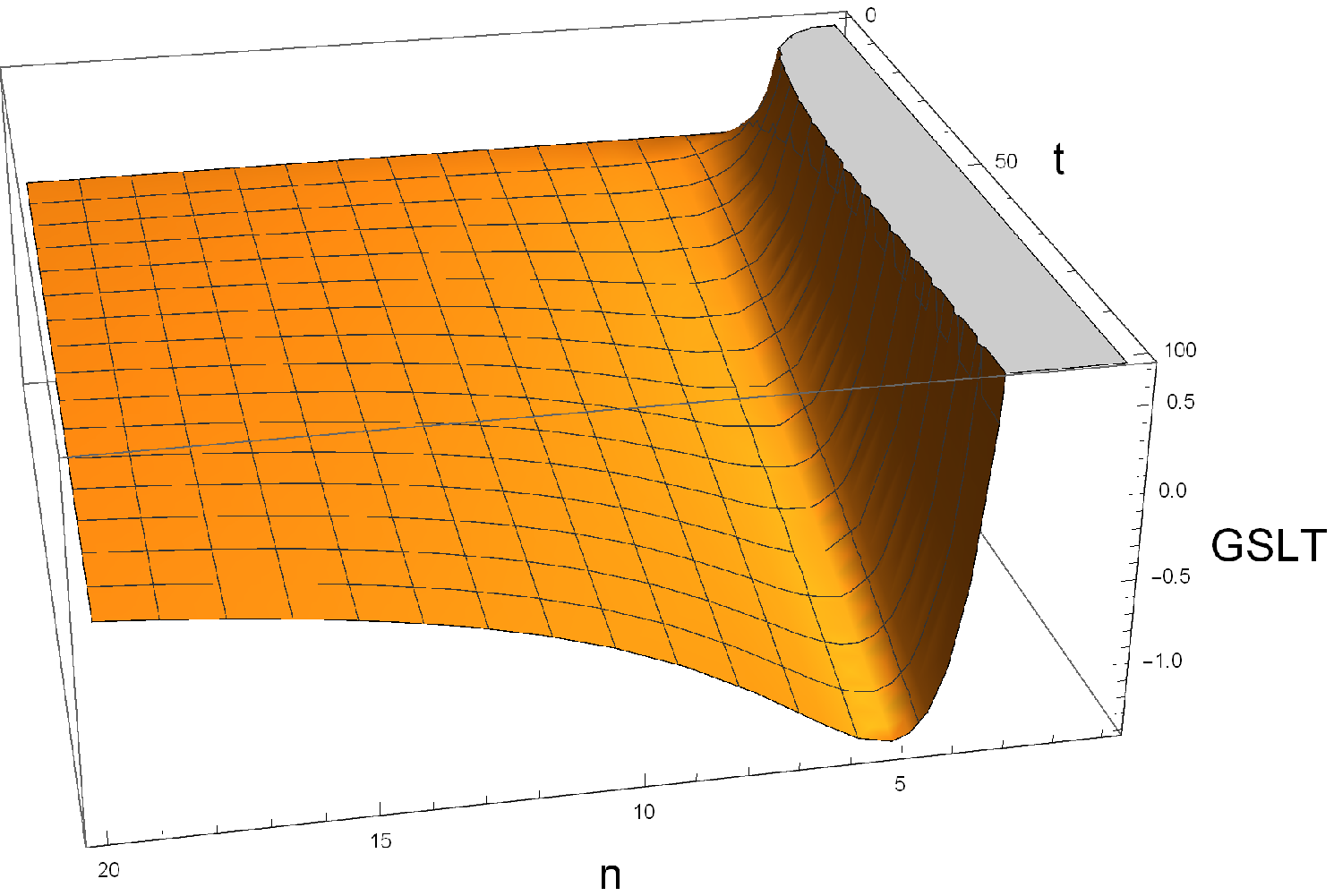}
    \includegraphics[width=0.4\textwidth]{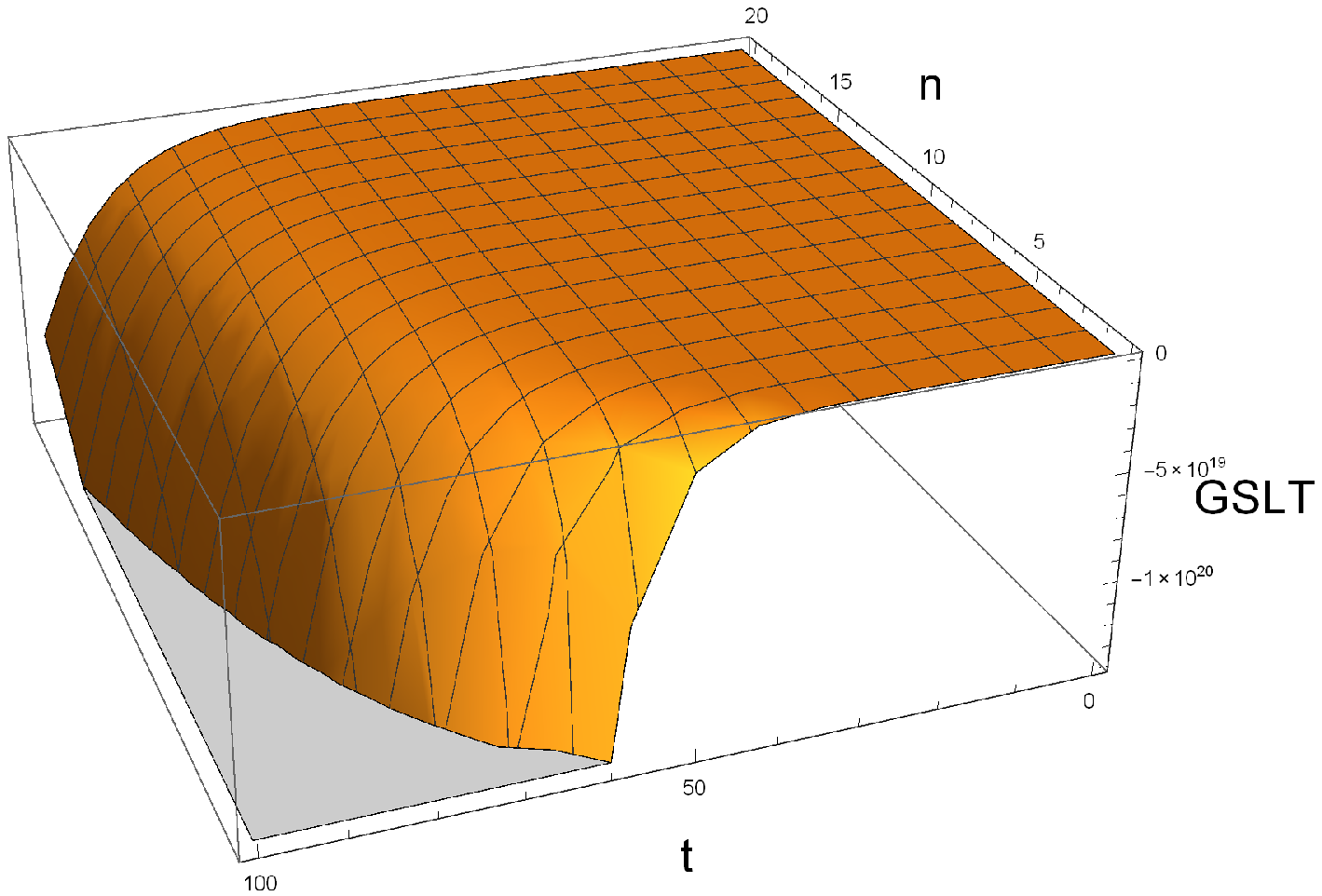}
    \caption{Validity of GSLT for exponential coupling for $\alpha=-0.1$. The figure on the right (left) shows the case where $\phi_{0}=10$ ($\phi_{0}=0.1$)}
    \label{figG1}
\end{figure}

\begin{figure}[H]
    \centering
    \includegraphics[width=0.4\textwidth]{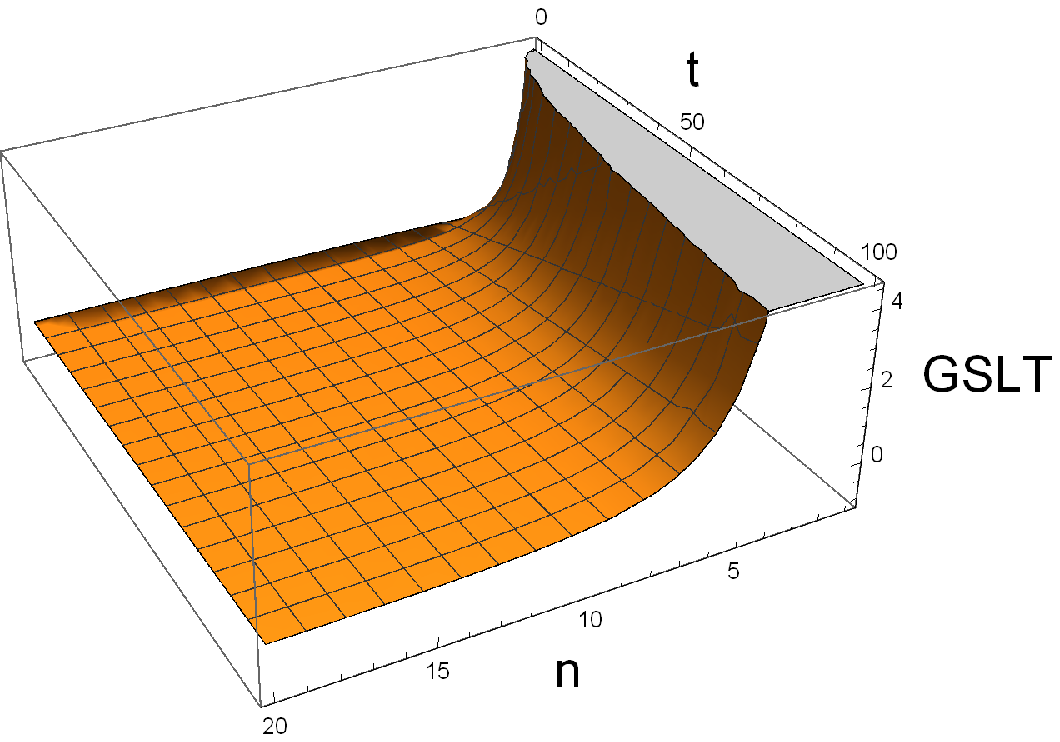}
    \includegraphics[width=0.4\textwidth]{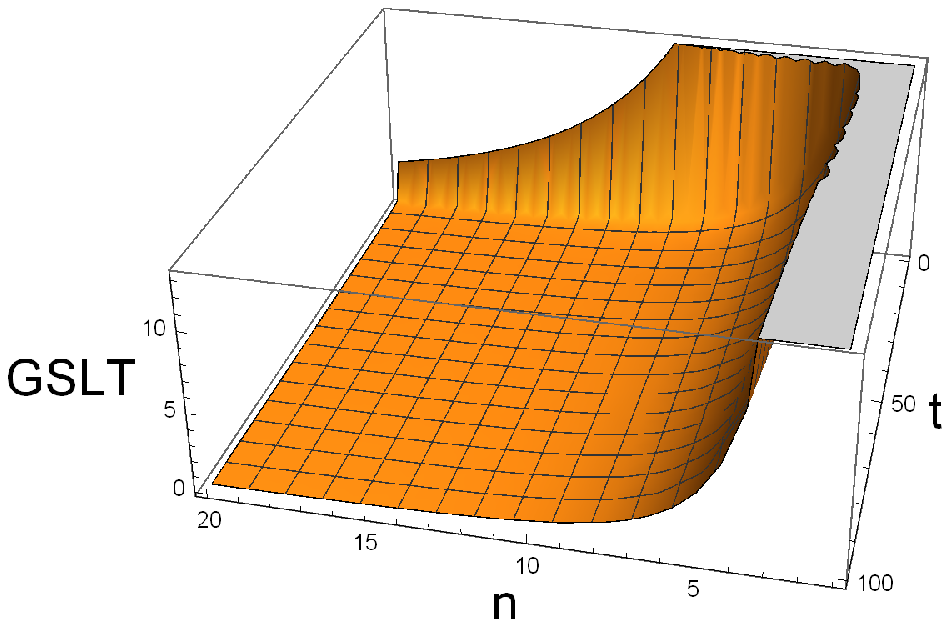}
    \caption{Validity of GSLT for exponential coupling for $\phi_{0}=-1.4$. The figure on the right (left) shows the case where $\alpha=-100$ ($\alpha=-0.1$)}
    \label{fig14}
\end{figure}

This section is devoted to analyse if the GSLT is satisfied for some exponential solutions found in Sec. \ref{exponential}. Since mostly all the solutions are similar, we will only analyse the case where $w=\tfrac{2-\phi_{0}}{3n}-1$ (see Sec.~\ref{case2}). In this case, the energy potential contains three exponentials that can represent different kind of potentials. For example, by taking $m=\tfrac{\sqrt{3} \kappa  \phi_0}{\sqrt{2(\phi_0-1)}}$, we find
\begin{eqnarray}
V(\phi)=\frac{(\phi_0-2) (\phi_0-1) e^{-\frac{\sqrt{6} \kappa  \phi }{\sqrt{\phi_0-1}}}}{3 \kappa ^2}-\frac{(\phi_0-1) \phi_0 (\alpha  (\phi_0-1)+3 \beta  (\phi_0-2)) e^{\frac{\sqrt{\frac{3}{2}} \kappa  \phi  (\phi_0-2)}{\sqrt{\phi_0-1}}}}{3 (\phi_0-2)}\,,
\end{eqnarray}
which is known as a double exponential potential. This kind of potential has been widely studied in the literature.
From Fig. \ref{Figg6}, it is observed that the scalar potential increases as the field increases. Thus the cosmic acceleration will be driven by the potential energy of the scalar field.\\
Another interesting example that can be constructed by taking $\phi_{0}=4$ and $\alpha=\tfrac{24 \kappa ^2-m^2}{48 \kappa ^4}$, which gives us the following potential
\begin{eqnarray}
V(\phi)=\frac{16 \left(m^2-24 \kappa ^2\right) \sinh \left(\frac{m \phi }{2}\right)}{m^4}-\frac{24 \beta  \kappa ^2 \left(24 \kappa ^2+m^2\right) e^{\frac{6 \kappa ^2 \phi }{m}-\frac{m \phi }{4}}}{m^4}\,,
\end{eqnarray}
which is an interesting kind of potential studied among the literature (see \cite{Sahni:1999gb,Matos:2000ss}).
The exponential function models generally lead to accelerating expansion behavior of the universe.
Let us know study the GSLT for this kind of potential. Under this model,  Eq. (\ref{29}) becomes
\begin{eqnarray}
\dot{S}_{tot}=\frac{\pi  t \left(\frac{2}{2 n-1}+\alpha  (\phi_0+2) t^{\phi_0}\right)}{n^2}\geq 0\,,\, \, \phi_{0}\neq2\,,
\end{eqnarray}
where for simplicity we took $\kappa=G=1$. Clearly, the above inequality will hold depending on the values of the parameters $\phi_{0},n$ and $\alpha$. Since $n>1$, if $\phi_{0}\geq-2$ and $\alpha>0$, GSLT will be valid at any time. Additionally, if $\phi_{0}\leq-2$ and $\alpha<0$, the GSLT will be always true. For all the other cases, the validity of GSLT will depend on time. Figs.~\ref{figG1} and \ref{fig14} show the behaviour of the GSLT inequality for different values of the parameters. For $\phi_{0}>0$ and $\alpha<0$, the inequality will be more constraint for bigger $|\phi_{0}|$. Additionally, it can be seen that GSLT will be always true at very late times for $-2<\phi_{0}<0$. .

\section{Generalized Second Law of Thermodynamics with Logarithmic entropy corrections}

The entropy area relation involving quantum corrections leads to
curvature corrections in the Einstein-Hilbert action \cite{21}. The
logarithmic corrected entropy is defined through the relation
\cite{22,23}
\begin{equation}\label{40}
S=\frac{Af(\phi)}{4G}+\alpha\ln\frac{Af(\phi)}{4G}+\beta\frac{4G}{Af(\phi)}+\gamma\,,
\end{equation}
where $\alpha$, $\beta$ and $\gamma$ are dimensionless constants,
however the exact values of these constants are yet to be
determined. These corrections arise in black hole entropy in loop
quantum gravity due to thermal equilibrium fluctuations and quantum
fluctuations \cite{24}. In \cite{23} Sadjadi and Jamil
investigated the validity of GSLT for FRW space-time with
logarithmic correction. They found that in a (super) accelerated
universe the GSL is valid whenever $\alpha(<)>0$ leading to a
(negative) positive contribution from logarithmic correction to the
entropy. In the following section, we will present the validity of the GSLT with modified entropy relations
involving logarithmic corrections.

First note that the time derivative of Eq. (\ref{40}) yields
\begin{equation}\label{40a}
\dot S=\Big( \frac{\dot A}{A}+\frac{f'}{f} \dot\phi  \Big)\frac{Af}{4G}\Big[ 1+\alpha\Big( \frac{4G}{Af} \Big) -\beta\Big( \frac{4G}{Af} \Big)^2 \Big]
\end{equation}
In case of the apparent horizon, Eq.~(\ref{40}) implies
\begin{equation}\label{41}
\dot{S}_A=\Big( \frac{-2\dot H}{H} +\frac{f'}{f} \dot\phi\Big)\frac{\pi f}{GH^2}\Big[  1+\alpha\Big(  \frac{GH^2}{f\pi} \Big) -\beta \Big(  \frac{GH^2}{f\pi} \Big)^2\Big]
\end{equation}
Here, we used that the Hubble horizon is $R_A=1/H$. Now, by
using Eq.~(\ref{27}) and (\ref{41}), we get
\begin{equation}\label{43}
\dot{S}_{tot}=\Big( \frac{-2\dot H}{H} +\frac{f'}{f} \dot\phi\Big)\frac{\pi f}{GH^2}\Big[  1+\alpha\Big(  \frac{GH^2}{f\pi} \Big) -\beta \Big(  \frac{GH^2}{f\pi} \Big)^2\Big]+\frac{4\pi}{G}\frac{\dot{H}(\dot{H}+H^2)}{(\dot{H}+2H^2)H^3}
\end{equation}
which can be represented in terms of effective components as follows
\begin{eqnarray}\label{44}
\dot{S}_{tot}&=&\Big(  \kappa   (w_\text{eff}+1)\sqrt{\frac{3\rho_\text{eff}}{f(\phi (t))}} +\frac{f'}{f} \dot\phi\Big)\frac{3 \pi  f(\phi (t))^2}{G \kappa ^2 \rho_\text{eff}}\Big[  1+\alpha\Big(  \frac{G \kappa ^2 \rho_\text{eff}}{3 \pi  f(\phi (t))^2} \Big) -\beta \Big( \frac{G \kappa ^2 \rho_\text{eff}}{3 \pi  f(\phi (t))^2} \Big)^2\Big]\nonumber\\
&&+\frac{6 \pi  (w_\text{eff}+1) (3 w_\text{eff}+1) }{G \kappa (1-3
w_\text{eff})}\sqrt{\frac{3f(\phi (t))}{\rho_\text{eff}}}\geq0\,.
\end{eqnarray}

The GSL with quantum corrections have natural implications in
the studies of the very early universe since quantum corrections are
directly linked with high energy and short distance scales.

\section{Conclusions}

The non-minimally coupling models in cosmology are frequently used
to study guaranteed late time acceleration; phantom crossing and the
existence of finite time future singularity. In this paper, the
thermodynamic study is executed in modified teleparallel theory
which involves scalar field non-minimally coupled to both the
torsion and a boundary term \cite{8}. 
One interesting factor in this theory is that under suitable limit, one can recover very
well-known theories of gravity as quintessence, teleparallel dark
energy and non-minimally coupled scalar field with the Ricci scalar $R$.

In this paper, we investigated the GSLT for an expanding universe
with apparent horizon. The laws of thermodynamics are universally valid
and hence must apply in a modified form to the whole universe. The
existence of a apparent horizon in a spacetime provides the opportunity to
formulate the first law  and the generalized second law of
thermodynamics. According to the laws of thermodynamics, the first
law must always be true for a physical system (provided it is
non-dissipative), however the GSLT holds exactly using the apparent
horizon.
We obtained these laws in the non-minimal coupled teleparallel theory
with a boundary term and scalar field. Our results generalize many previous GSLT
studies while extending them to the logarithmic corrected
entropy-area law. Moreover GSLT with quantum corrections have natural
implications in the studies of the very early universe since quantum
corrections are directly linked with high energy and short distance
scales.

We explored the existence of power law solutions for two different choices of Lagrangian coefficients,
which are functions of scalar field $f(\phi)$ and $g(\phi)$.
In first case we set $f(\phi)=1+\kappa^2\xi\phi^2$ and $g(\phi)=\kappa^2\chi\phi^2$ and
choose the power law forms of scale factor and scalar filed to discuss possible forms
of the power law potential $V(\phi)$ (see [28]). Here, One can recover the quadratic,
quartic, inverse and Ratra-Peebles potentials. We also discuss the Brans-Dicke theory
as a particular case in power law cosmology. Moreover, we also discuss the
power law solutions for the choice of $f(\phi)=1+\kappa^2\alpha e^{m\phi}$ and $g(\phi)=\kappa^2\beta e^{k\phi}$
with $\phi(t)=\frac{\phi_0}{m}\log(t)$. Here, we explored different form of $V(\phi)$ depending
on equation of state parameter.

We showed the behavior of matter density $\rho$ and potential $V(\phi)$ for power law potential,
it can be seen that both are decreasing functions of time as shown in Fig.\ref{figG4}.
In our discussion on validity of GSLT, we considered the quartic and inverse in all the
viable special cases like TDE, NMC-B, NMC-R and NMC-(B+R) and presented results in Figs.~\ref{figP1}-\ref{figP5}.
In Figs.~\ref{figP1} and \ref{figP2}, we find that GSLT is true for NMC-B (with $0<\xi<\frac{1}{6}$) and
TDE (with $0<\chi<\frac{1}{12}$). For inverse potential, we find that GSLT is valid for NMC-B (with $\frac{-1}{8}<\chi<0$) and
TDE (with $-\frac{2}{3}<\xi<0$), whereas it is violated for TDE as shown in Figs.~\ref{figP3} and \ref{figP4}.
For the choice of $\chi=\frac{1}{4}$, we found the validity region only in case of NMC-(B+R) with $n>\frac{1}{2}$
as shown in Fig.~\ref{figP5}. For exponential forms of Lagrangian coefficients, we found that GSLT is always true for
($\phi_0\leq-2$ \& $\alpha<0$) and ($\phi_0\geq-2$ \& $\alpha>0$) whereas for other choice of parameters validity of GSLT is time dependent.
Finally, we also formulated the GSLT for logarithmic entropy corrections.


\vspace{.5cm}
\acknowledgments
S.B. is supported by the Comisi{\'o}n Nacional de Investigaci{\'o}n
Cient{\'{\i}}fica y Tecnol{\'o}gica (Becas Chile Grant
No.~72150066). The authors would like to thank the anonymous referee for the interesting feedback and useful comments on the paper.

\appendix

\section{New cosmological solutions}
\label{appendix}
\subsection{Power-law solutions for $f(\phi)=1+\kappa^2\xi\phi^2$ and $g(\phi)=\kappa^2\chi\phi^2$}\label{powerlaw}
In this section we are going to find analytical power-law solutions for the NMC-(B+T) case where the coupling functions are
\begin{eqnarray}
f(\phi)&=&1+\kappa^2\xi\phi^2\,,\\
g(\phi)&=&\kappa^2\chi \phi^2\,,
\end{eqnarray}
where $\xi$ and $\chi$ are coupling constants. As we discussed before, for this specific choice we can recover NMC-B ($\xi=0$), TDE ($\chi=0$), NMC-R ($\xi=-\chi$) and quintessence theories ($\chi=\xi=0$).
Let us now study power-law cosmology where the scale function and the scalar field take a power-law form
\begin{eqnarray}
a(t)&=&a_{0}t^{n}\,, \ \ \ n>1\,,\\
\phi(t)&=&\phi_{0}t^{m}\,,
\end{eqnarray}
where $a_{0}$, $\phi_{0}$, $m$ and $n$ are constants. Under these ansatzes, from (\ref{FE1}) we directly find that the energy potential becomes
\begin{eqnarray}
V(\phi)&=&-\rho_{0}a_{0}^{-3 (w+1)} \phi ^{\frac{2-3 n (w+1)}{m}-\frac{2}{m}} \phi_{0}^{\frac{3 n (w+1)}{m}}-\frac{1}{2} \phi ^{2-\frac{2}{m}} \phi_{0}^{2/m} \left(m^2+12 m n \chi -6 n^2 \xi \right)\nonumber\\
&&+\frac{3 n^2 \phi ^{-2/m} \phi_{0}^{2/m}}{\kappa ^2}\,.\label{V1}
\end{eqnarray}
Now, by replacing the above expression in (\ref{FE2}) we find that the parameters need to satisfy
\begin{eqnarray}
\kappa ^2 \phi_{0}^{2/m} \phi  ^{2-2/m} \left[m^2 (4 \chi -1)-2 m (2 n \xi +3 n \chi +\chi )+2 n \xi \right]\nonumber\\
=
\kappa ^2 \rho_{0} (w+1) a_{0}^{-3 (w+1)} \phi_{0}^{\frac{3 n (w+1)}{m}} \phi ^{-\frac{3 n (w+1)}{m}}
+2 n \phi_{0}^{2/m} \phi ^{-2/m}\,.\label{ola}
\end{eqnarray}
Since $n\neq0$, the exponents of the scalar field of the first and second terms on the r.h.s. must match otherwise the above equation will not hold. Therefore, the state parameter needs to be
\begin{eqnarray}
w=\frac{2}{3 n}-1\,.
\end{eqnarray}
Using this condition, Eq. (\ref{ola}) becomes
\begin{eqnarray}
\frac{2 \phi_{0}^{2/m} \phi ^{-2/m} \left(\kappa ^2 \rho_{0}a_{0}^{-2/n}-3 n^2\right)}{3 n}&=&\kappa ^2 \phi_{0}^{2/m} \phi  ^{2-2/m} (m^2 (4 \chi -1)-2 m (2 n \xi \nonumber\\
&&+3 n \chi +\chi )+2 n \xi )\,.
\end{eqnarray}
This condition will be true for different choices of the parameters. For simplicity, we will assume the case that $a_{0}=1$, which gives us
\begin{eqnarray}
\rho_0&=&\frac{3 n^2}{\kappa ^2}\,,\\
m&=&\begin{cases}
\displaystyle\frac{2n(2 \xi +3\chi) +2 \chi \pm\sqrt{4 (2 n \xi +(3 n+1) \chi )^2+ 8 n \xi  (1-4 \chi)}}{2 (4 \chi -1)}\,,\ \ \ \chi\neq\frac{1}{4}\label{m12}\,,\\ \\
\displaystyle\frac{4 n \xi }{8 \xi  n+3 n+1}\,, \ \ \ \chi=\frac{1}{4}\,,n\neq- \frac{1}{8 \xi +3}\,.\label{m12b}
\end{cases}
\end{eqnarray}
Thus, the energy potential (\ref{V1}) and the energy density simplified to be
\small{\begin{eqnarray}
	\rho_{m}(t)&=&\frac{3n^{2}}{\kappa^2}t^{-2}\,,\\
	V(\phi)&=&\begin{cases}
	\-\displaystyle\frac{1}{2} \phi ^{2-2/m} \phi_{0}^{2/m} \left(m^2+12 m n \chi -6 n^2 \xi \right)\,, \chi\neq\frac{1}{4}\,, m=\tfrac{2n(2 \xi +3\chi) +2 \chi \pm\sqrt{4 (2 n \xi +(3 n+1) \chi )^2+ 8 n \xi  (1-4 \chi)}}{2 (4 \chi -1)} \,,\\ \\
	\displaystyle\frac{n^2 \xi  (8 \xi +3) \left(3 n^2 (8 \xi +3)-1\right) \phi ^{-\frac{3 n+1}{2 n \xi }-2}\phi_{0}^{\frac{3 n+1}{2 n \xi }+4}}{(n (8 \xi +3)+1)^2}\,, \ \ \ \chi=\frac{1}{4}\,,m=\frac{4 n \xi }{8 \xi  n+3 n+1}\,,n\neq- \frac{1}{8 \xi +3}\,.
	\end{cases}
	\end{eqnarray}}
For the first potential ($\chi\neq1/4$), we can recover a self-interacting scalar field case $V(\phi)\propto \phi^4$ if $m=-1$ which is the case where the coupling parameters satisfied $\xi=\tfrac{-6 n \chi -6 \chi +1}{6n}$. It corresponds to the inverse potential if $m=2/3$ with the additional constraint $\xi=-\frac{2(1-\chi+9n\chi)}{3n}$ and Ratra-Peebles potential is recovered by choosing $0 <m <1$.


\subsection{Power-law solutions for Brans-Dicke: $f(\phi)=\kappa^2\xi \phi^2$ and $g(\phi)=\kappa^2\chi \phi^2$}
It is easy to see that if we choose
\begin{eqnarray}
f(\phi)&=&\kappa^2\xi \phi^2\,,\\
g(\phi)&=&\kappa^2\chi \phi^2\,,
\end{eqnarray}
our action becomes a Brans-Dicke one in a canonical form with the Brans-Dicke parameter being $w_{BD}=1$. The later theory has been widely studied in the literature, so that it is an important coupling to consider. \\
To find solutions, we will follow the same approach as the previous section. Moreover, since the equations are so similar, we will not state all the steps. It is easy to see that if $w=-1+\frac{2 (1-m)}{3 n}\neq -1$ we obtain the following solutions,
\small{\begin{eqnarray}
	\rho_{m}(t)&=&
	\begin{cases}
	\frac{3 n\phi_{0}^2 t^{2 m-2} \left(m^2 (1-4 \chi )+2 m (2 n \xi +3 n \chi +\chi )-2 n \xi \right)}{2 (m-1)} & \rho_{0}=\frac{3 n\phi_{0}^2 a_{0}^{\frac{2}{n}-\frac{2 m}{n}} \left(-4 m^2 \chi +m^2+4 m n \xi +6 m n \chi +2 m \chi -2 n \xi \right)}{2 (m-1)}\,\\
	\rho_{0}& n=\frac{2 \chi -1}{2 (\xi +3 \chi )}\,, \ m=1\,.
	\end{cases}\,,\nonumber\\
	V(\phi)&=&
	\begin{cases}
	-\frac{m \phi ^{2-\frac{2}{m}}\phi_{0}^{2/m} \left(m^2+m (3 n-1)+6 n (n (\xi +3 \chi )-\chi )\right)}{2 (m-1)}\,, \ \ 
	\rho_{0}=\frac{3 n\phi_{0}^2 a_{0}^{\frac{2}{n}-\frac{2 m}{n}} \left(-4 m^2 \chi +m^2+4 m n \xi +6 m n \chi +2 m \chi -2 n \xi \right)}{2 (m-1)}\,, & \\
	\frac{-2 \xi ^2 \left(2\rho_{0}+\text{$\phi $0}^2\right)-3 \xi  \left(8\rho_{0} \chi +\left(4 \chi ^2+4 \chi -1\right)\phi_{0}^2\right)+18 \chi ^2 \left((1-4 \chi )\phi_{0}^2-2\rho_{0}\right)}{4 (\xi +3 \chi )^2}\,, 
	n=\frac{2 \chi -1}{2 (\xi +3 \chi )}\,, \ m=1\,. &
	\end{cases}\,.\nonumber
	\end{eqnarray}}

\subsection{Exponential solutions for $f(\phi)=1+\kappa^2\alpha e^{m\phi}$ and $g(\phi)=\kappa^2\beta e^{k\phi}$}\label{exponential}
In this section, we will assume that the coupling functions are exponential as follows
\begin{eqnarray}
f(\phi)&=&1+\kappa^2\alpha e^{m\phi}\,,\\
g(\phi)&=&\kappa^2\beta e^{k\phi}\,,
\end{eqnarray}
where $\alpha$, $\beta$, $m$ and $k$ are parameters. Cosmology for the specific case where $\beta=-\alpha$ and $k=m$ was studied in \cite{Pettorino:2004zt}.
Additionally, we assume that the scale factor and the scalar field behave as
\begin{eqnarray}
a(t)&=&a_{0}t^{n}\,, \ \ \ n>1\,,\\
\phi(t)&=&\frac{\phi_{0}}{m}\log(t)\,,
\end{eqnarray}
where $n,m,\phi_{0}$ and $a_{0}$ are constants. By using (\ref{FE1}), we find that the energy potential becomes
\begin{eqnarray}
V(\phi)&=&-\rho_{0} a_{0}^{-3 (w+1)}e^{-\frac{3 m n (w+1) \phi }{\phi_{0}}}-e^{-\frac{2 m \phi }{\phi_{0}}} \left(\frac{3 \beta  k n \phi_{0} e^{k \phi }}{m}+\frac{\phi_{0}^2}{2 m^2}-\frac{3 n^2}{\kappa ^2}\right)+3 \alpha  n^2 e^{\frac{m \phi  (\phi_{0}-2)}{\phi_{0}}} \,.
\end{eqnarray}
Now, if we replace this equation in (\ref{FE2}), we find that that
\begin{eqnarray}\label{cond}
2 \alpha  \kappa ^2 m^2 n (\phi_{0}-1) a_{0}^{3 w+3} e^{\frac{m \phi  (3 n (w+1)+\phi_{0})}{\phi_{0}}}+a_{0}^{3 w+3} \left(\kappa ^2 \phi_{0}^2-2 m^2 n\right) e^{\frac{3 m n (w+1) \phi }{\phi_{0}}}&=&\nonumber \\
\beta  \kappa ^2 k \phi_{0} a_{0}^{3 w+3} (k \phi_{0}-m (3 n+1)) e^{\phi  \left(k+\frac{3 m n (w+1)}{\phi_{0}}\right)}-\kappa ^2 m^2 \rho_{0} (w+1) e^{\frac{2 m \phi }{\phi_{0}}}&&
\end{eqnarray}
needs to hold. For the specific case where $\rho_{0}\neq0$, we have three possibles cases for the state parameter
\begin{eqnarray}
w=\begin{cases}
\displaystyle\frac{2 m-k \phi_0}{3 m n}-1\,,  \\ \\
\displaystyle\frac{2-\phi_0}{3 n}-1\,,\\ \\
-1
\end{cases}\,.
\end{eqnarray}
Now, we will explore all the possible solutions under these state
parameters. In general, all the solutions are very similar but they
are not exactly the same.

\subsubsection{Case 1: $w=\tfrac{2 m-k \phi_0}{3 m n}-1$}
For this special case, from (\ref{cond}) the other parameters must satisfy ($k\neq 2m$)
\begin{eqnarray}
\rho_{0}&=&-\frac{3 \beta  \kappa ^2 k \left(3 \kappa ^2-2 k m+2 m^2\right) a_{0}^{\frac{2 m (2 m-k)}{\kappa ^2}}}{4 m^4 (2 m-k)}\,, \ \  \phi_{0}= 1\,,n= \frac{\kappa ^2}{2 m^2}\,.
\end{eqnarray}
Note, that for the case $k=2m$ we find that that $n=1/3$ which is not allowed. Hence, if $k\neq 2m$ we find that the energy density and energy potential for the first state parameter $w_{1}$, are given by
\begin{eqnarray}
\rho_{m}(t)&=&\frac{\beta  k \left(-3 \kappa ^2+2 k m-2 m^2\right) \left(3 \kappa ^2+2 k m-4 m^2\right) t^{\frac{k}{m}-2}}{4 m^4 (k-2 m)}\,,\\
V(\phi)&=&\frac{3 \beta  k \left(2 \kappa ^2 m^2-3 \kappa ^4\right) e^{k \phi -2 m \phi }}{4 m^4 (k-2 m)}+\frac{3 \alpha  \kappa ^4 e^{-m \phi }}{4 m^4}+\frac{\left(3 \kappa ^2-2 m^2\right) e^{-2 m \phi }}{4 m^4}\,.
\end{eqnarray}
The special case non-minimally coupled with the Ricci scalar can be obtained by taking $k=m$ and $\beta=-\alpha$, which makes that the potential becomes
\begin{eqnarray}
V(\phi)&=&\frac{3 \beta  \kappa ^2 \left(\kappa ^2-m^2\right) e^{-m \phi }}{2 m^4}+\frac{\left(3 \kappa ^2-2 m^2\right) e^{-2 m \phi }}{4 m^4}\,.
\end{eqnarray}

\subsubsection{Case 2: $w=\tfrac{2-\phi_0}{3 n}-1$}\label{case2}
For this case, we need to have $\phi_0\neq 2$ and also from (\ref{cond}) the parameters must obey
\begin{eqnarray}
\rho_0=\frac{3 \alpha  \kappa ^4 (\phi_0-1) \phi_0^4 a_{0}^{\frac{4 m^2}{\kappa ^2 \phi_0^2}-\frac{2 m^2}{\kappa ^2 \phi_0}}}{2 m^4 (\phi_0-2)}\,,\ k=\frac{3 m n+m}{\phi_0}\,,\  n=\frac{\kappa ^2 \phi_0^2}{2 m^2}\,,
\end{eqnarray}
and hence the energy density and the potential take the following form
\begin{eqnarray}
\rho_m (t)&=&\frac{3 \alpha  \kappa ^4 (\phi_0-1) \phi_0^4 t^{\phi_0-2}}{2 m^4 (\phi_0-2)}\,,\\
V(\phi)&=&-\frac{3 \alpha  \kappa ^4 \phi_0^5 e^{m \phi -\frac{2 m \phi }{\phi_0}}}{4 m^4 (\phi_0-2)}-\frac{3 \beta  \left(3 \kappa ^4 \phi_0^4+2 \kappa ^2 m^2 \phi_0^2\right) e^{\frac{3 \kappa ^2 \phi  \phi_0}{2 m}-\frac{m \phi }{\phi_0}}}{4 m^4}\nonumber\\
&&+\frac{\left(3 \kappa ^2 \phi_0^4-2 m^2 \phi_0^2\right) e^{-\frac{2 m \phi }{\phi_0}}}{4 m^4}\,.
\end{eqnarray}
\subsubsection{Case 3: $w= -1$ (dark energy fluid)}
For a dark-energy fluid with $w=-1$, from  (\ref{cond}) the parameters need to be
\begin{eqnarray}
\phi_0= 1\,, \ k= 3 m n+m\,, \ n= \frac{\kappa ^2}{2 m^2}\,,
\end{eqnarray}
and therefore the energy density is a constant $\rho_{0}$ and the potential becomes
\begin{eqnarray}
V(\phi)&=&\frac{3 \alpha  \kappa ^4 e^{-m \phi }}{4 m^4}-\frac{3 \beta  \left(3 \kappa ^4+2 \kappa ^2 m^2\right) e^{\phi  \left(\frac{3 \kappa ^2}{2 m}+m\right)-2 m \phi }}{4 m^4}+\frac{\left(3 \kappa ^2-2 m^2\right) e^{-2 m \phi }}{4 m^4}-\rho_{0}\,.
\end{eqnarray}

\end{document}